 \documentclass[final,3p,times]{elsarticle}

\usepackage{multirow,setspace,amssymb,amsmath,graphicx,color,rotating,subfigure,url,lineno,natbib}
\usepackage{textcomp}

\journal{Physica A} 

\begin{document}

\begin{frontmatter}

\title{Statistical properties of online avatar numbers in a
massive multiplayer online role-playing game}
\author[BS,SS,RCE]{Zhi-Qiang Jiang}
\author[BS,RCE,RCSE]{Fei Ren}
\author[BS,SS,RCE]{Gao-Feng Gu}
\author[SNDA]{Qun-Zhao Tan}
\author[BS,SS,RCE,RCSE,RCFE]{Wei-Xing Zhou \corref{cor}}
\cortext[cor]{Corresponding author. Address: 130 Meilong Road, P.O.
Box 114, School of Business, East China University of Science and
Technology, Shanghai 200237, China, Phone: +86 21 64253634, Fax: +86
21 64253152.}
\ead{wxzhou@ecust.edu.cn} %

\address[BS]{School of Business, East China University of Science and Technology, Shanghai 200237, China}
\address[SS]{School of Science, East China University of Science and Technology, Shanghai 200237, China}
\address[RCE]{Research Center for Econophysics, East China University of Science and Technology, Shanghai 200237, China}
\address[RCSE]{Engineering Research Center of Process Systems Engineering (Ministry of Education), East China University of Science and Technology, Shanghai 200237, China}
\address[SNDA]{Shanda Interactive Entertainment Ltd, Shanghai 201203, China}
\address[RCFE]{Research Center on Fictitious Economics \& Data Science, Chinese Academy of Sciences, Beijing 100080, China}

\begin{abstract}
Massive multiplayer online role-playing games (MMORPGs) are very
popular in past few years. The profit of an MMORPG company is
proportional to how many users registered, and the instant number of
online avatars is a key factor to assess how popular an MMORPG is.
We use the on-off-line logs on an MMORPG server to reconstruct the
instant number of online avatars per second and investigate its
statistical properties. We find that the online avatar number
exhibits one-day periodic behavior and clear intraday pattern, the
fluctuation distribution of the online avatar numbers has a
leptokurtic non-Gaussian shape with power-law tails, and the
increments of online avatar numbers after removing the intraday
pattern are uncorrelated and the associated absolute values have
long-term correlation. In addition, both time series exhibit
multifractal nature.
\end{abstract}


\end{frontmatter}

\section{Introduction}
\label{S1:Intro}

A massive multiplayer online role-playing game (MMORPG) is a genre
of online role-playing games (ORPGs) in which a large number of
players interact with one another within a virtual world. The term
MMORPG was coined by Richard Garriott, who created Ultima Online. In
mainland China, Shanda Interactive Entertainment Ltd is the leader
of the MMORPG industry, which is based in Shanghai. Shanda runs
dozens of online games and has most registered players.

An MMORPG forms an online virtual world, where people can work and
interact with one another in a somewhat realistic manner. Therefore,
virtual worlds have great potential for research in the social,
behavioral, and economic sciences \cite{Bainbridge-2007-Science}.
For instance, we can design a kind of virus in a virtual world and
let it spread to investigate its epidemics, we can design some
economic games in a virtual world to study the formation of human
cooperation (indeed, numerical experiments have been done
\cite{Grabowski-Kosinski-2008-APPA}), and we can record the economic
behaviors of avatars to understand the evolution of wealth
distribution. A pioneering work was done by Edward Castronova, who
traveled in a virtual world called ``Norrath'' and performed
preliminary analysis of its economy \cite{Castronova-2001-WP}.
Recently, there are also efforts in the field of computational
social sciences from a complex network perspective
\cite{Grabowski-2007-PA,Grabowski-Kruszewska-2007-IJMPC,Grabowski-Kruszewska-Kosinski-2008-EPJB,Grabowski-Kosinski-2008-APPB,Grabowski-Kruszewska-Kosinski-2008-PRE,Grabowski-2009-PA}.
In addition to its scientific potentials, virtual worlds could act
as nice places for real social activities, such as marketing
\cite{Matsuda-2003-Presence,Castronova-2005-HBR,Hemp-2006-HBR}, and
provide opportunities for players to make real money
\cite{Papagiannidis-Bourlakis-Li-2008-TFSC}.

In this work, we investigate the behavior of instant online avatar
numbers in a server of a very popular MMORPG. The number of instant
online avatars is of crucial importance for scientific and
commercial purposes. The paper is organized as follows. In Section
\ref{S1:Data}, we describe the data and the procedure to construct
the time series of instant online avatar numbers. Section
\ref{S1:Seasonality} studies the seasonal patterns of the time
series. The probability distribution of the fluctuations of online
avatar numbers is researched in Section \ref{S1:PDF}, and the
temporal correlations and multifractal properties are analyzed in
Section \ref{S1:Memory}. Section \ref{S1:Conclusion} summarizes the
main findings of this paper.

\section{Data preprocessing}
\label{S1:Data}

The MMORPG game investigated is called ``Legend of Mir'', which is
copyrighted and run by Shanda Interactive Entertainment Ltd in
Shanghai. This online game was very popular in China several years
ago.

\begin{figure}[htb]
\centering
\includegraphics[width=8cm]{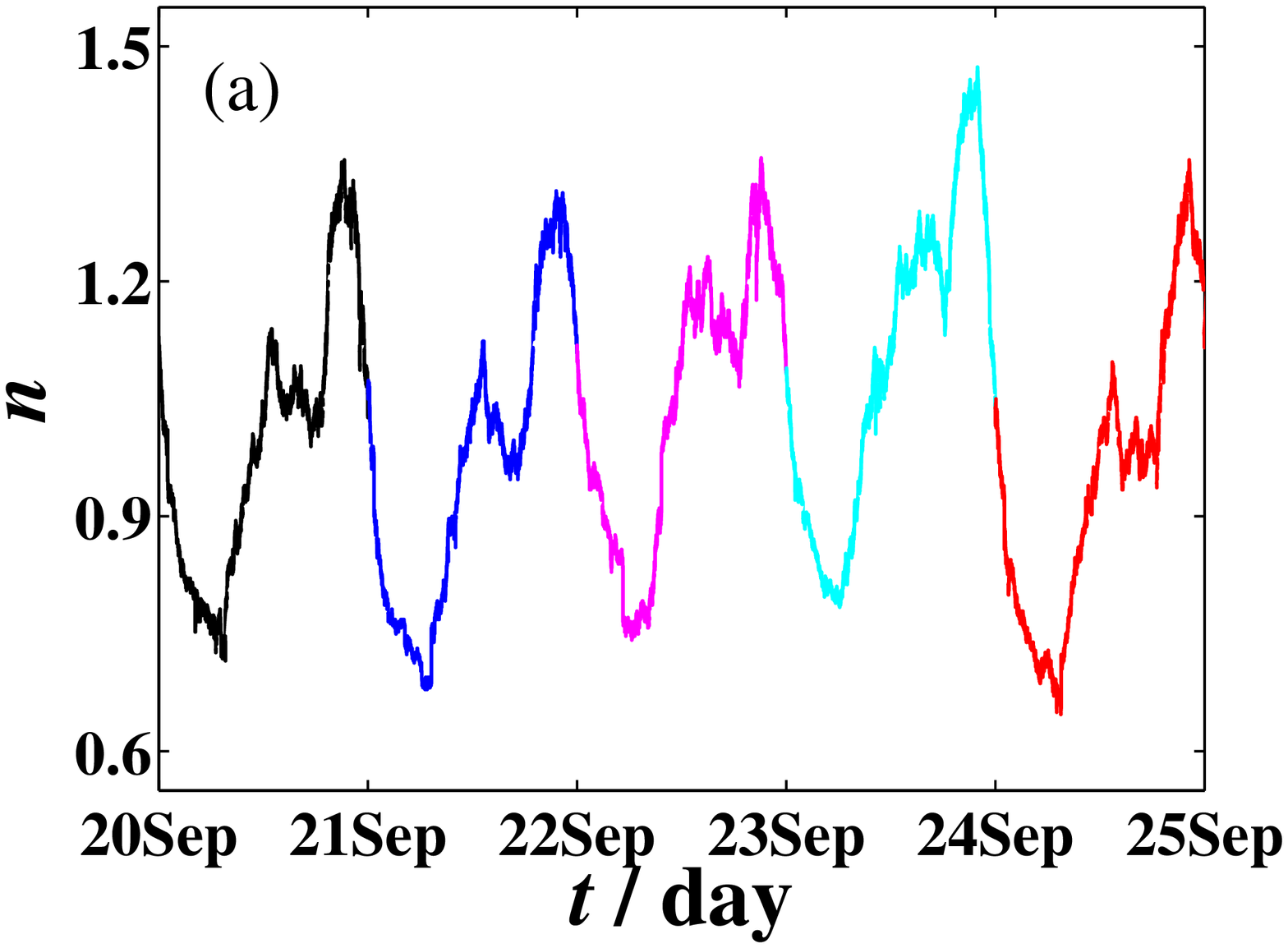}
\includegraphics[width=8cm]{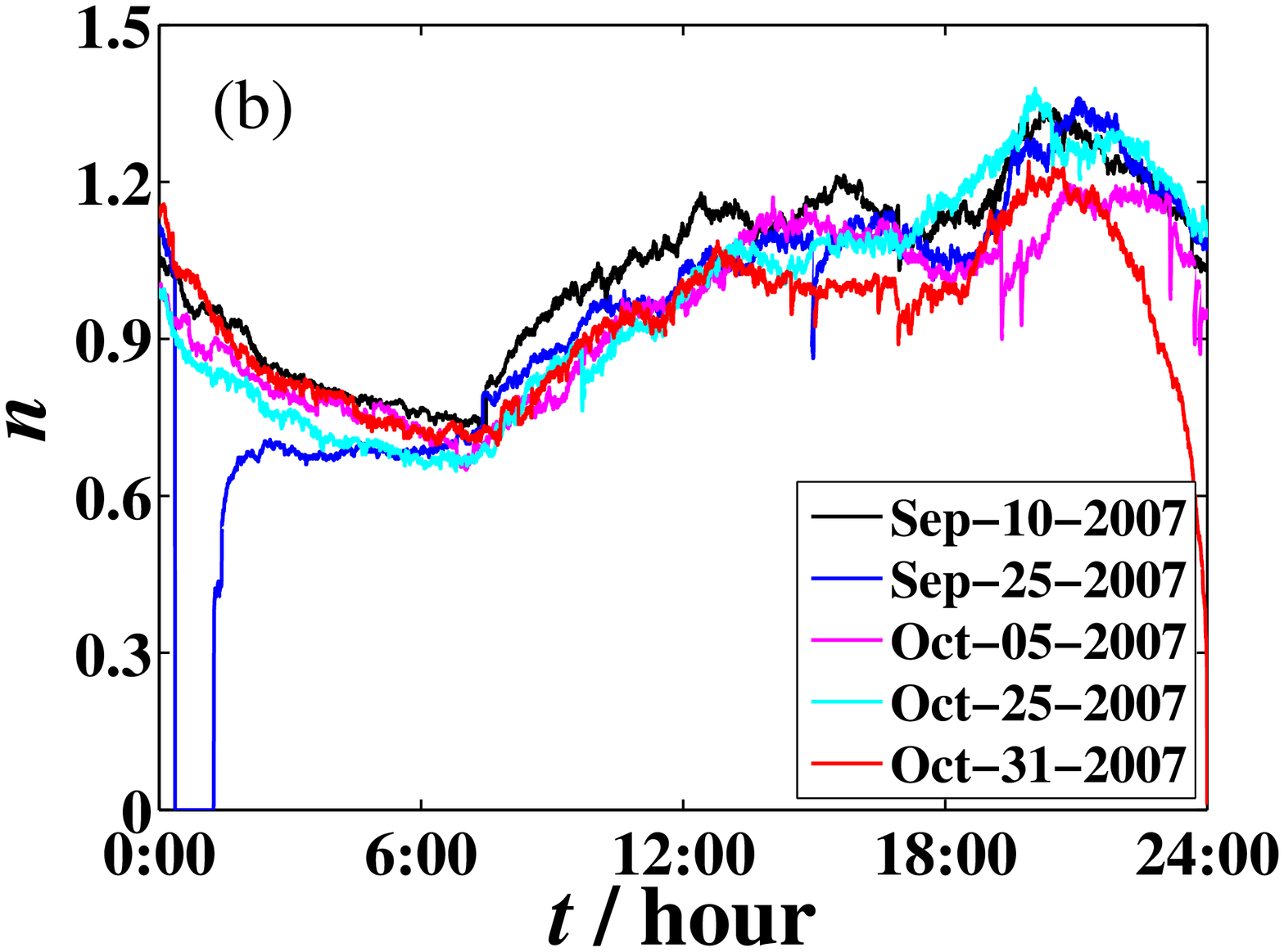}
\caption{\label{Fig:NoaExample} (color online) The online avatar
numbers $n(t)$ for (a) a continuous period of five days and (b) five
separate days.}
\end{figure}

An avatar is activated when a player logs on an MMORPG server. When
he quits the game, the server records an entry including the time
moments of his log-on and log-off, accurate to one second. An
on-off-line log is saved at the end of each day. This allows us to
reconstruct the number of online avatars $n(t)$ at each second $t$.
The data we analyzed are from 1 September 2007 to 31 October 2007.
The variable is divided by its mean in the presentation of this
paper, which does not change the results. A segment of online avatar
numbers is plotted in Fig.~\ref{Fig:NoaExample} (a). It is not
surprising that there is a daily periodic pattern in the evolution
of online avatar number $n(t)$. The number $n(t)$ has an intraday
low at around 7:30 and then increases in the following daytime, with
a plateau from 12:00 to 18:00. At around 20:00, $n(t)$ reaches its
maximum value. This pattern repeats every day.

Fig.~\ref{Fig:NoaExample}(b) illustrates the plots of online avatar
numbers on five days including 10 September 2007, 25 September 2007,
5 October 2007, 25 October 2007, and 31 October 2007. One finds that
the curves almost share the same shape except for 25 September 2007
and 31 October 2007. For 25 September 2007, the online avatar number
fell rapidly and remained zero between 00:30 and 01:30 in the
morning. This phenomenon is observed for several days in the sample,
which is due to the fact that the server was scheduled for
maintaining or game version updating after midnight. However,
Fig.~\ref{Fig:NoaExample} suggests that the maintaining time is
better to be around 7:00 (say, from 6:30 to 7:30) in the morning in
order to impact less players. For 31 October 2007, there was a sharp
decease of the online avatar number at the end of the day. This
reflects a finite-size effect or a boundary effect. Note that our
data set is truncated at 23:59:59 on 31 October 2007 and the
on-off-line logs are recorded based on the log-off time. This means
that the logs of 31 October 2007 exclude the situation that the
avatar went online before 23:59:59 and offline after 23:59:59.
Therefore, the online avatar numbers in this last day are excluded
from our analysis, and the resultant data set has 60 days.

\section{Seasonal patterns}
\label{S1:Seasonality}

\subsection{Spectral analysis}
\label{S2:FFT}

\begin{figure}[htb]
\centering
\includegraphics[width=8cm]{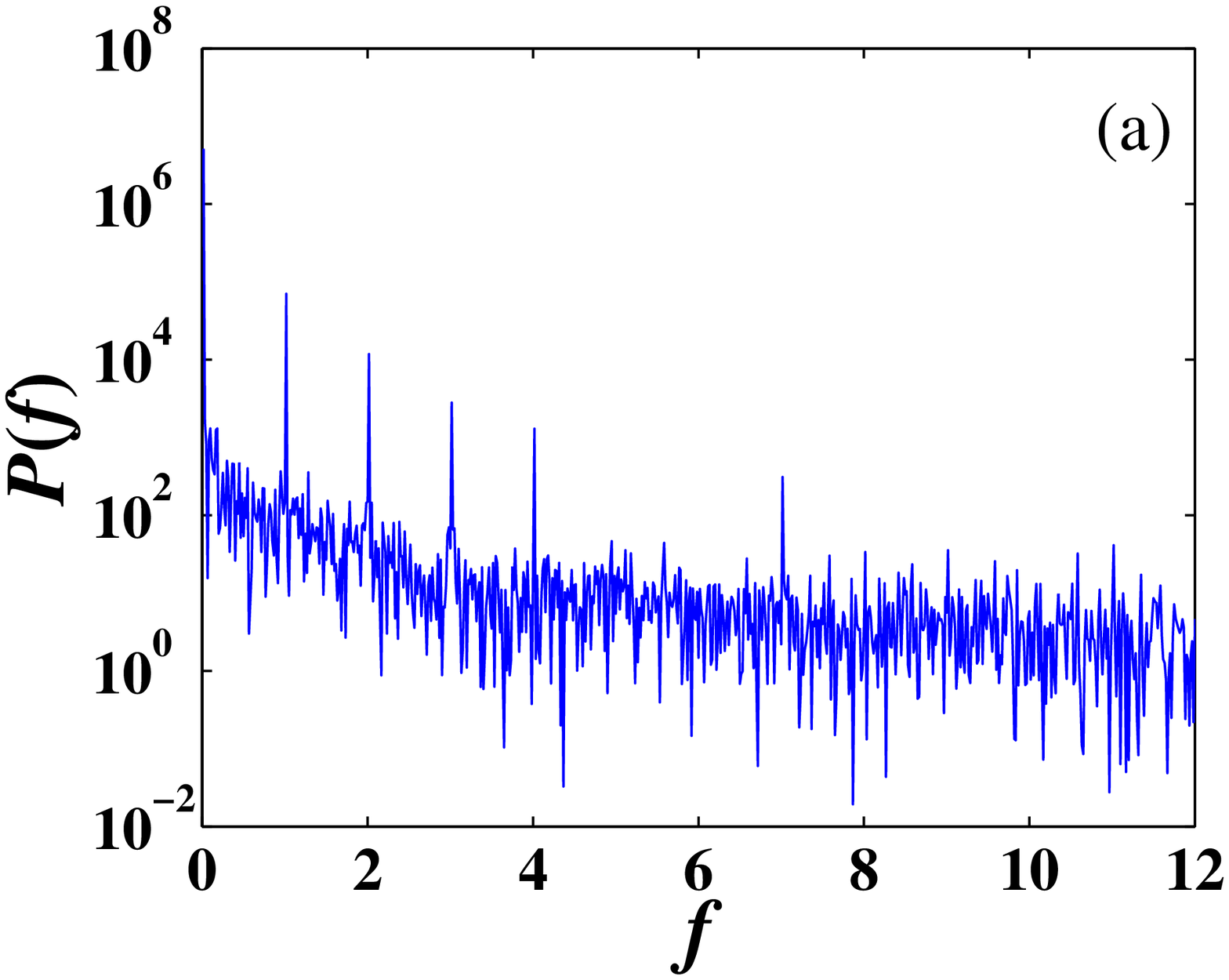}
\includegraphics[width=8cm]{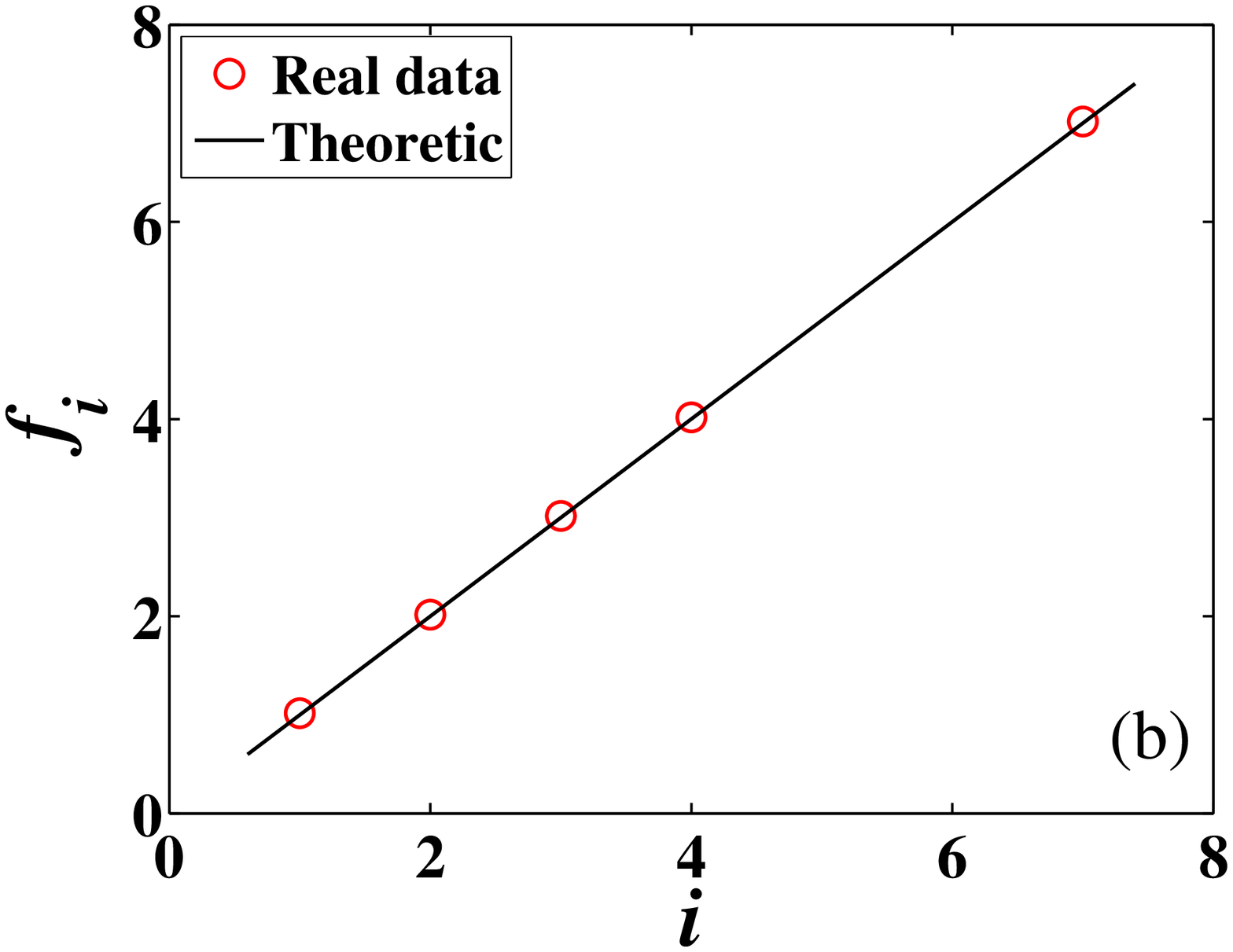}
\caption{\label{Fig:FFT} Spectral analysis of online avatar numbers
by means of fast Fourier transformation: (a) Power spectrum; (b)
Fundamental frequency.}
\end{figure}

Because of the circadian and weekly cycles of human activity,
evident periodicity is observed in the time series of online avatar
numbers. A spectral analysis is adopted to quantify the periodic
behavior. Fig.~\ref{Fig:FFT}(a) illustrates the power spectrum of
online avatar number series. The units of the frequency $f$ is
1/day. The highest peak lies at $f=0.0167$, which captures nothing
but the weak global trend of the time series
\cite{Zhou-Sornette-2002-IJMPC}. The second highest peak locates at
$f_1=1.0167$, which is statistically significant with a $p$-value of
$3.41\times10^{-47}$
\cite{Press-Teukolsky-Vetterling-Flannery-1996}. It implies that the
periodicity is about one day, as suggested by
Fig.~\ref{Fig:NoaExample}(a). We also see harmonic peaks around
$f=2,3,...$, which further confirms that the observed one-day
periodicity is not artificial.

In order to determine the fundamental frequency $f_0$ using harmonic
peaks, we can regress the following equation between the harmonic
frequencies $f_i$ and the corresponding orders of the associated
peaks,
\cite{Zhou-Sornette-2002-PD,Zhou-Sornette-Pisarenko-2003-IJMPC,Zhou-Sornette-2009-PA}.
\begin{equation}
 f_i = a + i \times f_0.
 \label{Eq:fi}
\end{equation}
We choose the peaks that are significant higher than its neighboring
spectral powers in Fig.~\ref{Fig:FFT}(a), and five peaks at
$f_1=1.0167$, $f_2=2.0167$, $f_3=3.0167$, $f_4=4.0167$, and
$f_7=7.0167$ are determined. The plot of $f_i$ against $i$ shown in
Fig.~\ref{Fig:FFT}(b) exhibits nice linearity. A linear
least-squares regression gives $a=0.0167$ and $f_0=1.000$. The
F-test finds the relation (\ref{Eq:fi}) significant with a $p$-value
of zero. The t-test shows that the coefficient $a$ is significantly
different from zero and the hypothesis $f_0=1$ can not be rejected
at the significance level 0.0001. This fundamental frequency $f_0=1$
corresponds to an exact one-day periodicity, which is the base for
the search of a possible intraday pattern.

\subsection{Intraday pattern and weekly pattern}
\label{S2:patterns}

In order to investigate the seasonal patterns in the time series, we
define the average online avatar number, calculated as follows,
\begin{equation}
\langle n(t) \rangle = \frac{1}{\#\cal{D}} \sum_{j\in\cal{D}}n_j(t),
 \label{Eq:ip}
\end{equation}
where $\cal{D}$ stands for the set of $\#\cal{D}$ days under
consideration and $n_j(t)$ represents the online avatar numbers on
$j$-th day of $\cal{D}$. For instance, if we need to estimate the
intraday average online avatar numbers for all the holidays,
$\cal{D}$ is the set containing all the holidays (Saturday, Sunday,
and public holidays) in the two corresponding months. Note that the
periods that have vanishing online avatar numbers are excluded from
the averaging procedure.

\begin{figure}[htb]
\centering
\includegraphics[width=6cm]{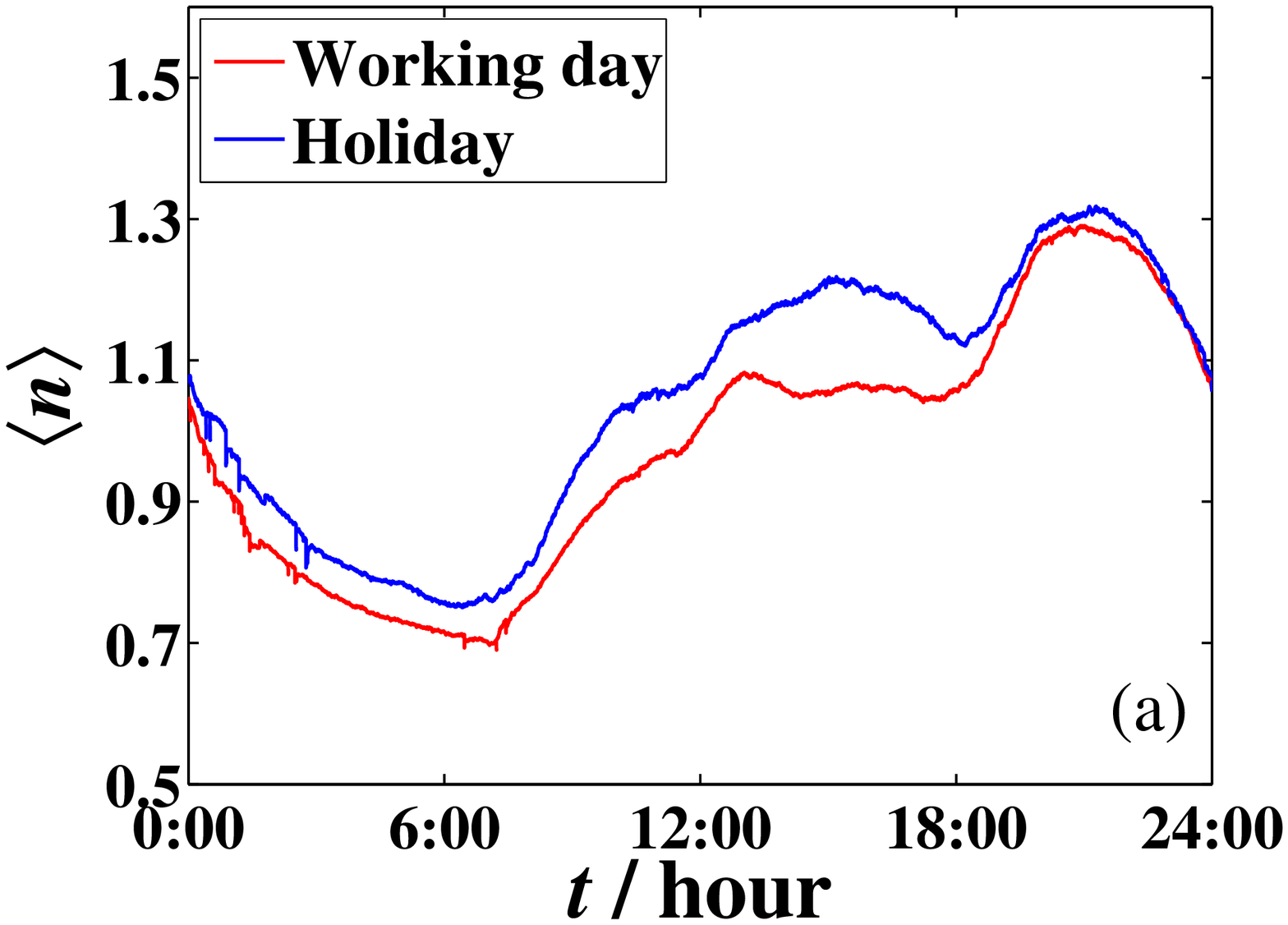}
\includegraphics[width=6cm]{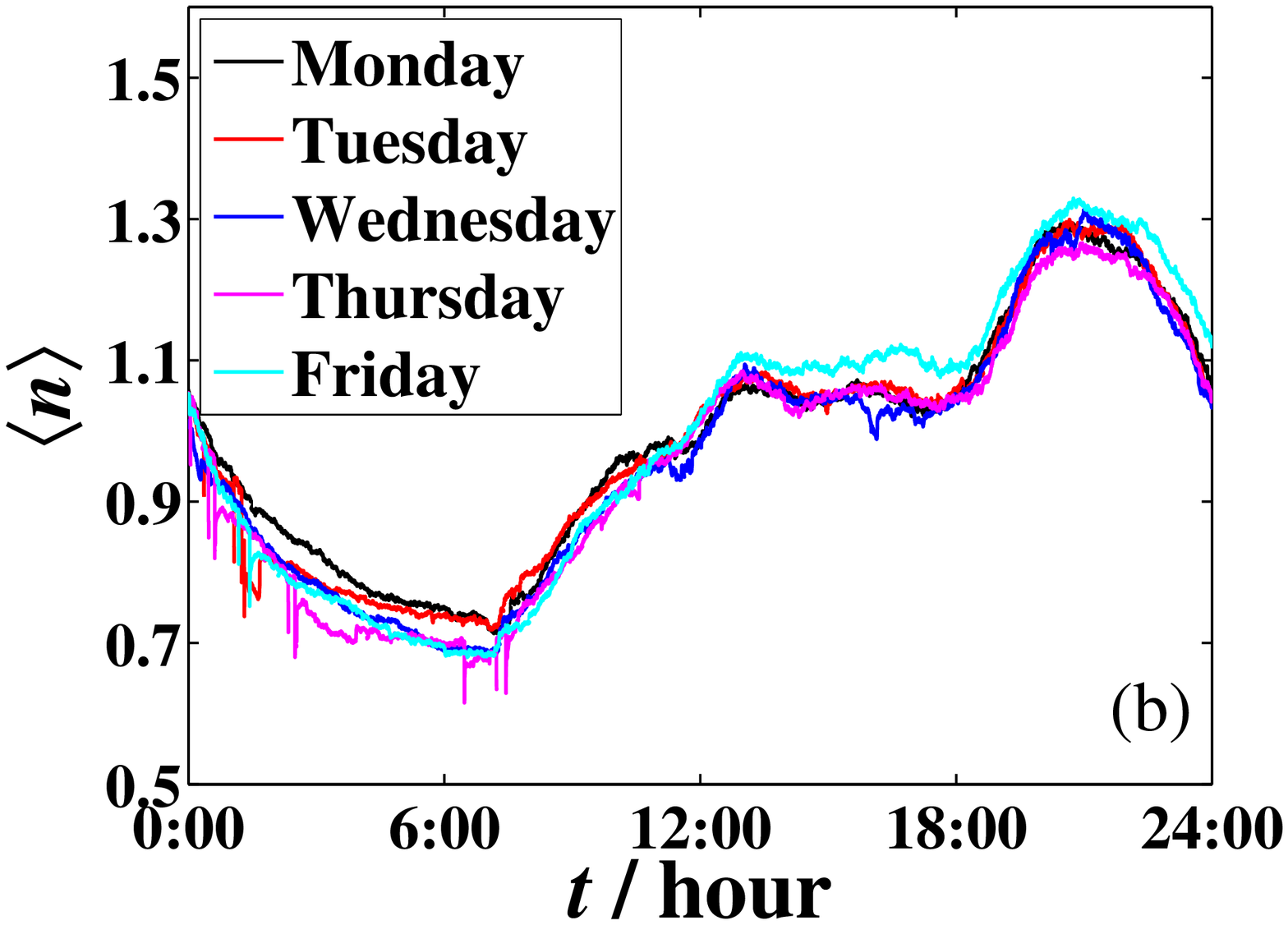}
\includegraphics[width=6cm]{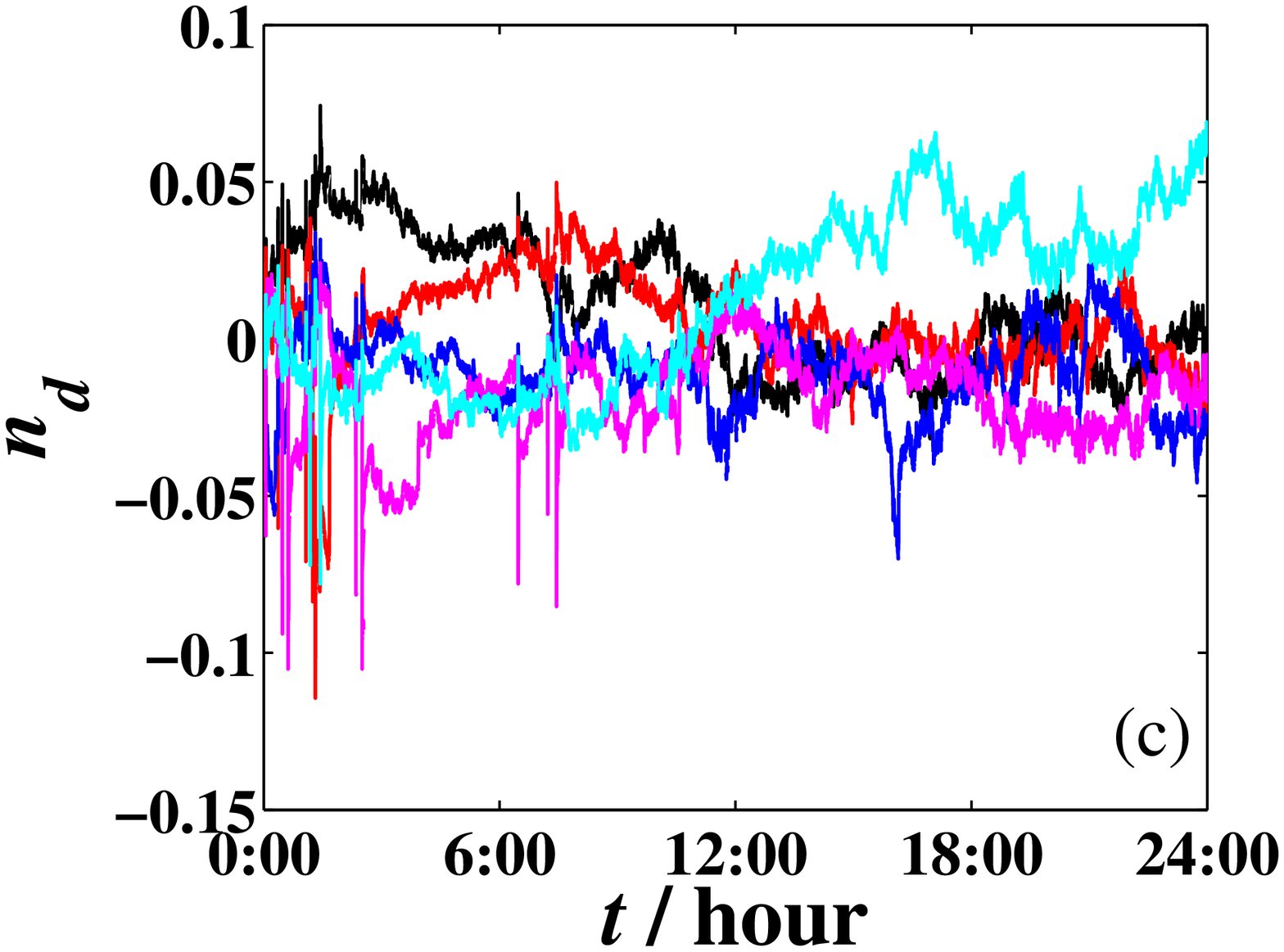}
\includegraphics[width=6cm]{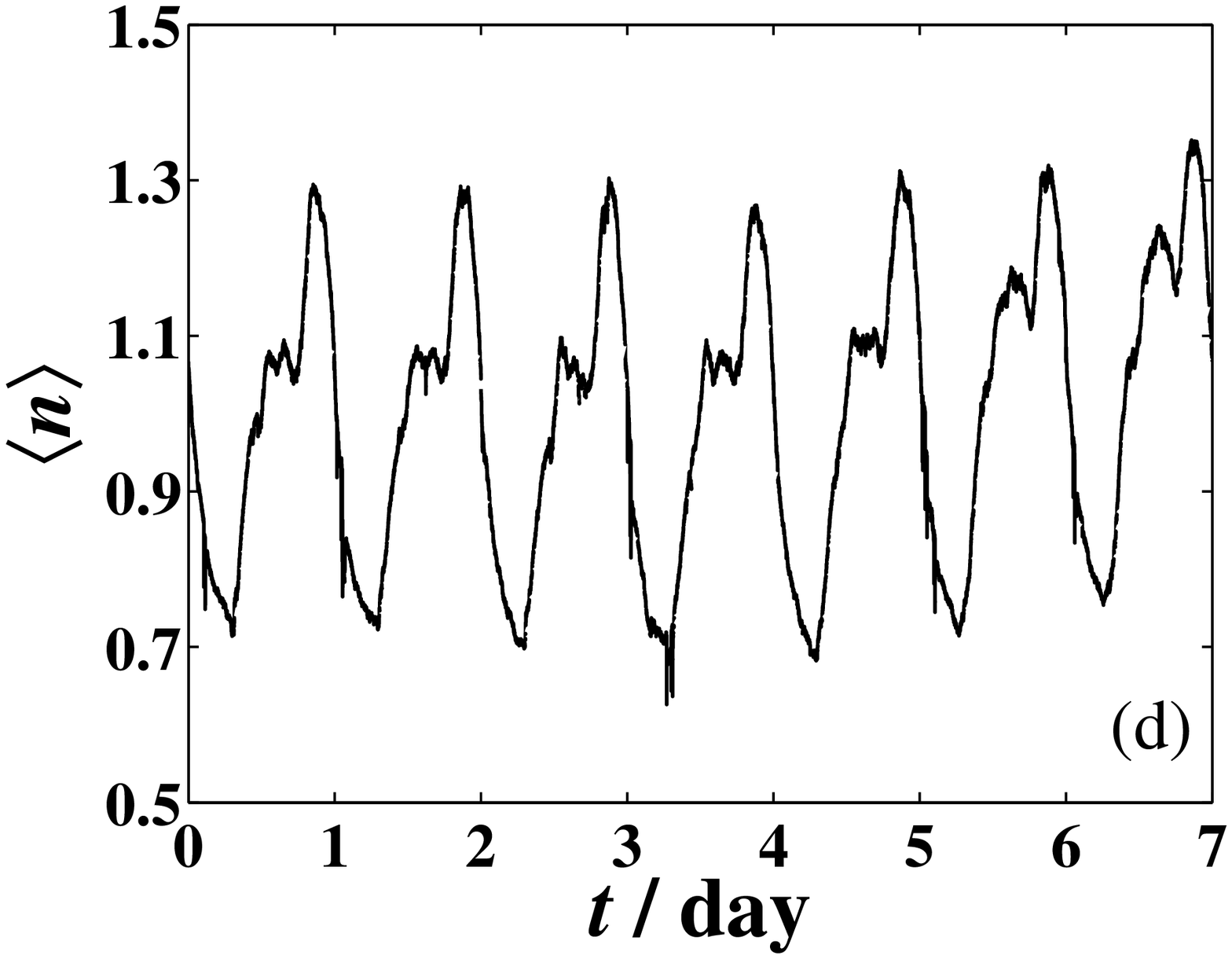}
\caption{\label{Fig:Pattern} Seasonal patterns of online avatar
numbers: (a) Intraday pattern for working days, and holidays
(including Saturday, Sunday, and public holidays); (b) Intraday
pattern for Monday, Tuesday, Wednesday, Thursday, and Friday
(consider only the sample of working day); (c) Dependence of $n_d$
as a function of $t$; (d) Weekly pattern.}
\end{figure}

As a first step, we partition all 60 days into two groups, one
containing all working days and the other including all weekends and
public holidays. The intraday patterns of these two groups of days
are shown in Fig.~\ref{Fig:Pattern}(a). The average online number
grows gradually after 7:30 in the morning. During the time period
between 12:00 and 18:00, the online number is almost stationary.
After 18:00, the online number increases again before around 20:00.
Then, the number drops till 7:30 in the next day. The trend of
online avatar number is consistent with the circadian rhythm of
human activities. In addition, the average online avatar number for
holidays is larger than that for working days, which is trivial. For
working days, we also calculate the intraday patterns for Monday,
Tuesday, Wednesday, Thursdays, and Friday, which are presented in
Fig.~\ref{Fig:Pattern}(b). Roughly speaking, the five curves almost
overlap and no remarkable difference is revealed among these days.
However, a careful scrutiny unveils that the average number curve
for Friday keeps above other days from noon to midnight. The pattern
in Friday evening is explained by the fact that Friday is followed
by Saturday and most of the players are free on Saturday, while that
in the Friday afternoon is explained by the fact that most college
students do not have courses and many official institutions have
much less work to do, for instance, only a small part of the
officials might have meetings. This Friday afternoon pattern is
expected to be idiosyncratic for MMORPGs played mainly by Chinese.

In order to further distinguish the differences among the intraday
patterns of Monday, Tuesday, Wednesday, Thursday, and Friday, we
define that
\begin{equation}
n_d(t) = \langle n_i(t) \rangle - \langle n_w(t) \rangle,
 \label{Eq:nd}
\end{equation}
where $i = 1,2,3,4,5$ stand for the five kinds of days (Monday,
Tuesday, Wednesday, Thursday, and Friday), $\langle n_i(t) \rangle$
is the intraday pattern of each kind of weekdays, and $\langle
n_w(t) \rangle$ represents the intraday pattern of working days
shown in \ref{Fig:Pattern}(a). Fig.~\ref{Fig:Pattern}(c) depicts the
dependence of $n_d$ as a function of $t$. One can observe that the
online numbers of Mondays (Fridays) are much larger than those of
the four other days before (after) 12:00. This suggests a weak
weekly pattern, which is illustrated in Fig.~\ref{Fig:Pattern}(d).
However, this weekly pattern is very weak and considering only the
intraday pattern is sufficient for most quantitative analyses.

\section{Probability distributions of $\Delta n$}
\label{S1:PDF}

\begin{figure}[htb]
\centering
\includegraphics[width=5.3cm]{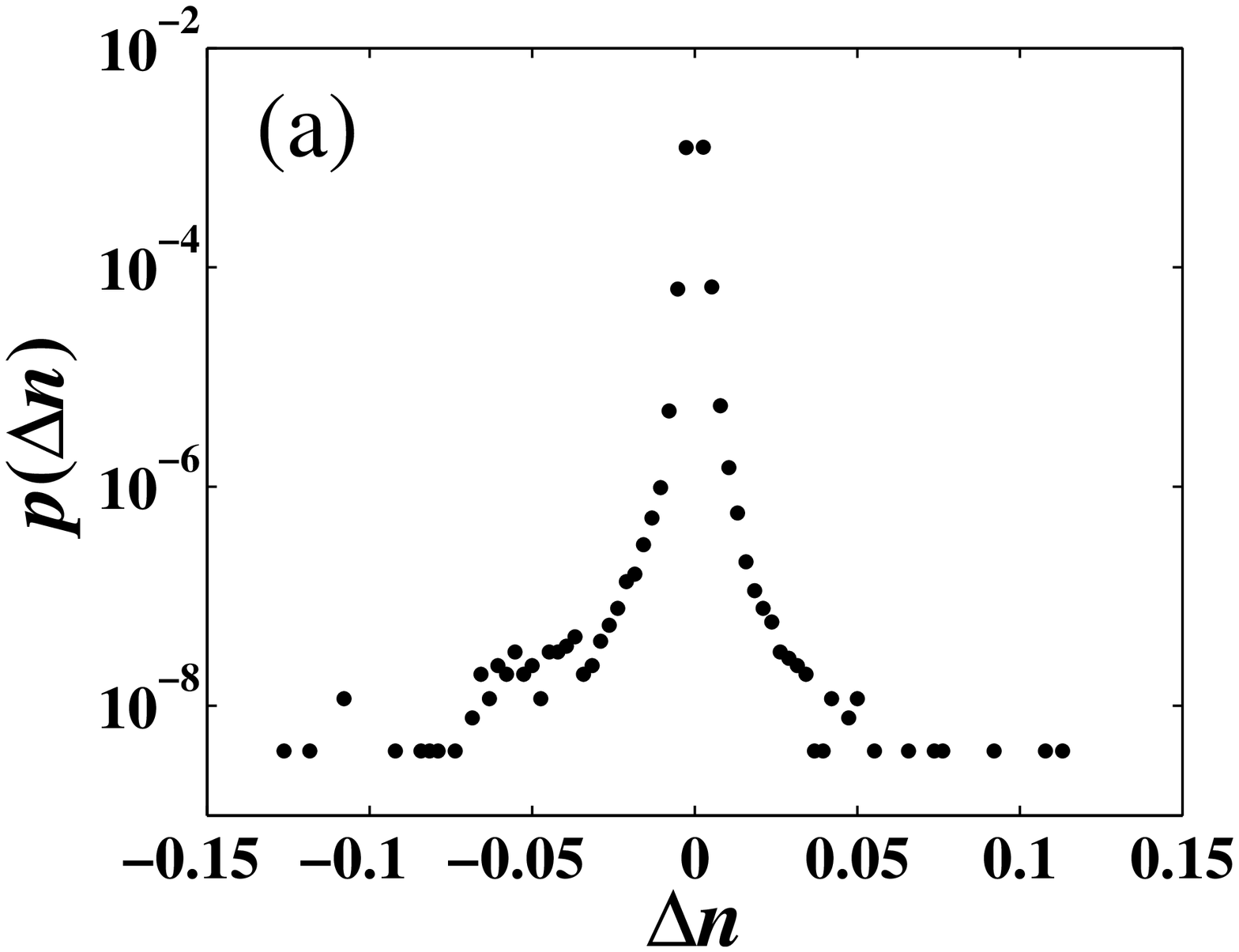}
\includegraphics[width=5.3cm]{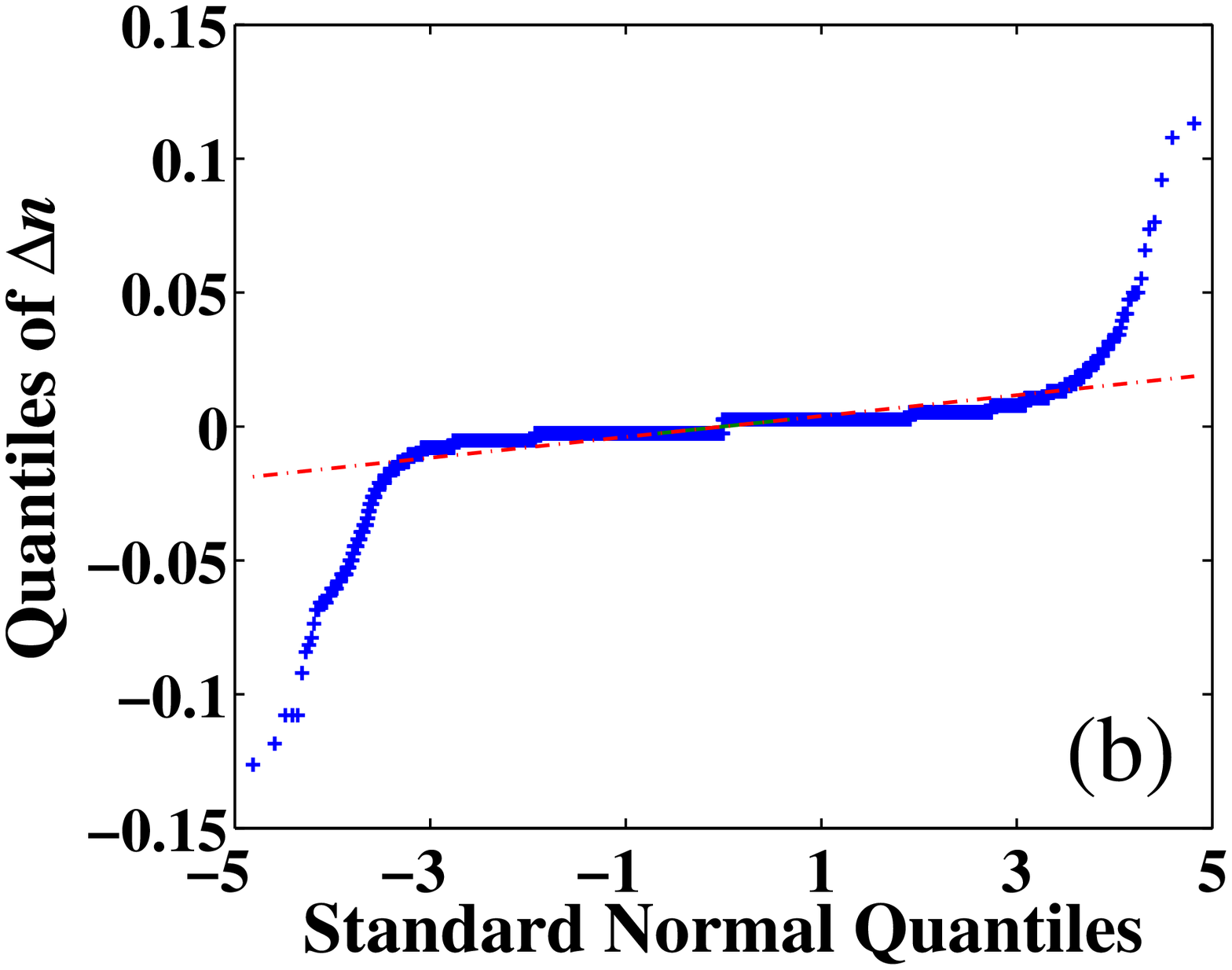}
\includegraphics[width=5.3cm]{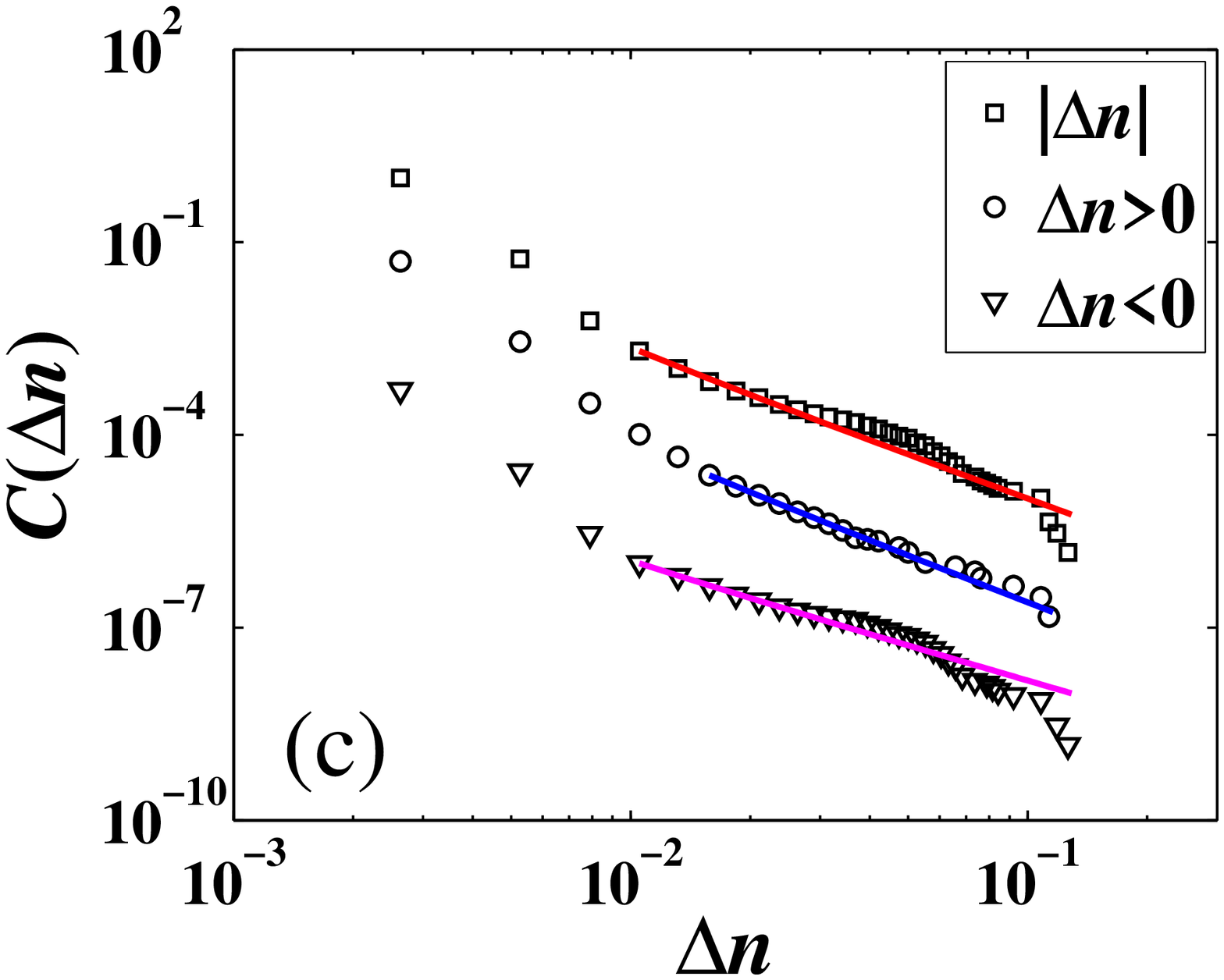}
\caption{\label{Fig:dn:PDF} (a) Probability density of the
fluctuations of online avatar numbers $\Delta n$. (b) QQ plot of
$\Delta n$. (c) Survival distribution of $\Delta n$. The curves of
$\Delta n
>0$ and $\Delta n <0$ have been translated vertically by a factor of
0.1 and 0.001 in turn for better visibility.}
\end{figure}

We now study the probability distribution of the fluctuations of
online avatar numbers $\Delta n$, which is defined as the difference
of online avatar numbers $n(t)$ in two successive seconds,
\begin{equation}
 \Delta n(t) = n(t) - n(t-1).
 \label{Eq:dn}
\end{equation}
Fig.~\ref{Fig:dn:PDF}(a) illustrates the empirical density function
of $\Delta n$ in log-linear scales. It is seen by eyeballing that
the distribution has a leptokurtic fat-tailed non-Gaussian shape.
The non-Gaussian feature of the fluctuation distribution can be
characterized by the QQ-plot shown in Fig.~\ref{Fig:dn:PDF}(b). To
investigate the tails, we show in Fig.~\ref{Fig:dn:PDF}(c) the
survival function of $|\Delta n|$, $\Delta n
>0$, and $\Delta n <0$. The method proposed by Clauset, Shalizi and Newman
\cite{Clauset-Shalizi-Newman-2009-SIAMR} has been employed to
confirm that the distributions have power-law tails,
\begin{equation}
 C(x) \sim x^{-\gamma}
 \label{Eq:C:x}
\end{equation}
when $x\geqslant x_{\min}$. We find that $\gamma= 3.24$ and
$x_{\min}=4$ for $x=|\Delta n|$, $\gamma= 3.38$ and $x_{\min}=6$ for
$x=\Delta n>0$,  and $\gamma=2.78$ and $x_{\min}=4$ for $x=\Delta
n<0$.

\section{Long memory and multifractality}
\label{S1:Memory}

In this section, we study the memory effect in the online avatar
number time series $n(t)$. As shown in Section \ref{S1:Seasonality},
there is an evident intraday pattern in the time series. This
periodic pattern is removed before correlation analysis for each day
$j$, which results in a new time series:
\begin{equation}
 n_r(t) = n_j(t) / \langle n \rangle,~~~~j=1,2,\cdots,60.
 \label{Eq:nr}
\end{equation}
The fluctuation of $n_r(t)$ is defined as follows
\begin{equation}
 \Delta n_r(t) = n_r(t) - n_r(t-1).
 \label{Eq:dnr}
\end{equation}
We stress that, $\Delta n_r(t)$ is the right quantity to check the
memory effect of $n(t)$, but investigating its distribution is of no
interest and seems meaningless.

\subsection{Long memory}
\label{S2:DFA}

The detrended fluctuation analysis (DFA) is utilized, which has the
ability to extract long-range power-law correlation in
non-stationary time series
\cite{Peng-Buldyrev-Havlin-Simons-Stanley-Goldberger-1994-PRE,Kantelhardt-Bunde-Rego-Havlin-Bunde-2001-PA}.
For a given intertrade duration series $\{\Delta n_r(t)|t =
1,2,\cdots,T\}$, we can define the cumulative summation series
$y(t)$ as follows,
\begin{equation}
 y(t) = \sum_{t'=1}^{t} \Delta n_r(t'),~~t = 1,2,\cdots,T.
  \label{Eq:cumsum}
\end{equation}
Note that the mean of the $\Delta n_r(t)$ series is not removed
before the cumulative summation, which will be removed in the
detrending step. The series $y(t)$ is covered by $N_s$ disjoint
boxes with the same size $s$. When the whole series $y(t)$ cannot be
completely covered by $N_s$ boxes, we can utilize $2N_s$ boxes to
cover the series starting from both ends of the time series. In each
box, a cubic polynomial trend function $g$ of the sub-series is
determined. The local detrended fluctuation function $f_k(s)$ in the
$k$-th box is defined as the r.m.s. of the fitting residuals:
\begin{equation}
 [f_k(s)]^2 = \frac{1}{s}\sum_{t=(k-1)s+1}^{ks} [y(t)-g(t)]^2~.
  \label{Eq:fk:s}
\end{equation}
The overall detrended fluctuation is estimated as follows
\begin{equation}
 [F_2(s)]^2 = \frac{1}{N_s}\sum_{k=1}^{N_s} \left[f_k(s)\right]^2.
  \label{Eq:F2:s}
\end{equation}
As the box size $s$ varies in the range of $[100,T/4]$, one can
determine the power-law relationship between the overall fluctuation
function $F_2(s)$ and the box size $s$, which reads,
\begin{equation}
 F_2(s) \sim s^H,
  \label{Eq:Hurst}
\end{equation}
where $H$ signifies the Hurst index, which is related to the power
spectrum exponent $\eta$ by $\eta = 2H-1$
\cite{Talkner-Weber-2000-PRE,Heneghan-McDarby-2000-PRE} and to the
autocorrelation exponent $\gamma$ by $\gamma = 2-2H$
\cite{Taqqu-Teverovsky-Willinger-1995-Fractals,Kantelhardt-Bunde-Rego-Havlin-Bunde-2001-PA}.

\begin{figure}[htb]
\centering
\includegraphics[width=8cm]{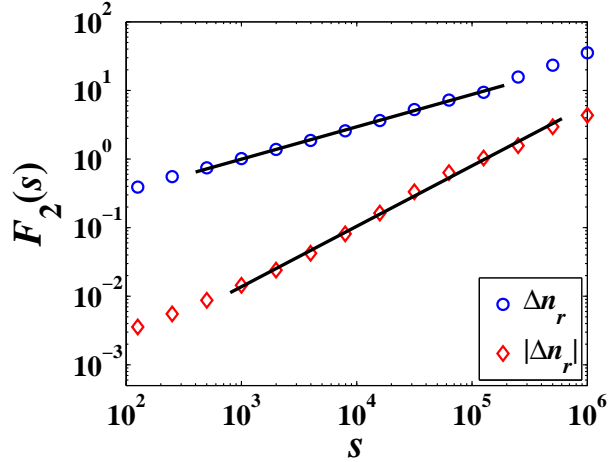}
\caption{\label{Fig:dnr:LM} (color online) Detrended fluctuation
analysis of $\Delta n_r(t)$ and $|\Delta n_r(t)|$. The solid lines
are the best power-law fits to data in the corresponding scaling
ranges. The hurst indexes are $H= 0.481 \pm 0.005$ for $\Delta n_r$
and $H = 0.868 \pm 0.012$ for $|\Delta n_r|$. }
\end{figure}

Fig.~\ref{Fig:dnr:LM} illustrates the dependence of the overall
fluctuation function $F_2(s)$ of $\Delta n_r(t)$ with respect to the
box size $s$ in double logarithmic coordinates. There is a nice
power-law relation spanning more than two orders of magnitude. A
linear least-squares regression of $\ln F_2(s)$ against $\ln s$
gives the estimate of the Hurst index $H = 0.481 \pm 0.005$. This
value is very close to $H=0.5$ and we argue that there is no
temporal correlation in the time series of $\Delta n_r(t)$. We also
show the DFA results for $|\Delta n_r(t)|$ in Fig.~\ref{Fig:dnr:LM}.
Again, we see a nice power law and the exponent is $H = 0.868 \pm
0.012$. In other words, there is evident long memory in the absolute
fluctuations $|\Delta n_r(t)|$ of online avatar numbers. This
observation is reminiscent of the behavior of stock returns
\cite{Mantegna-Stanley-2000}.

\subsection{Multifractal nature}

The DFA method can be extended to detect multifractal nature, known
as the multifractal detrended fluctuation analysis (MF-DFA)
\cite{Kantelhardt-Zschiegner-Bunde-Havlin-Bunde-Stanley-2002-PA}.
The overall detrended fluctuation in Eq.~(\ref{Eq:F2:s}) is
generalized to the following form
\begin{equation}
 F_q(s) = \left\{\frac{1}{N_s}\sum_{k=1}^{N_s}[f_k(s)]^q
 \right\}^{1/q}~,
  \label{Eq:Fq}
\end{equation}
where $q$ can take any real number except $q = 0$. When $q = 0$, we
have
\begin{equation}
 F_0(s) = \exp\left\{\frac{1}{N_s}\sum_{k=1}^{N_s}\ln[f_k(s)] \right\}.
  \label{Eq:F0}
\end{equation}
By varying the value of $s$ in the range from $s_{\min} = 100$ to
$s_{\max} = T/4$, one can expect the detrended fluctuation function
$F_q(s)$ scales with the size $s$:
\begin{equation}
 F_q(s) \sim s^{h(q)},
  \label{Eq:scaling}
\end{equation}
where $h(q)$ is the generalized Hurst index. Note that when $q = 2$,
$h(2)$ is nothing but the Hurst index $H$. We focus on $q \in [-4,
6]$ to obtain reasonable statistics in the estimation of $F_q(s)$.

\begin{figure}[htb]
\centering
\includegraphics[width=8cm]{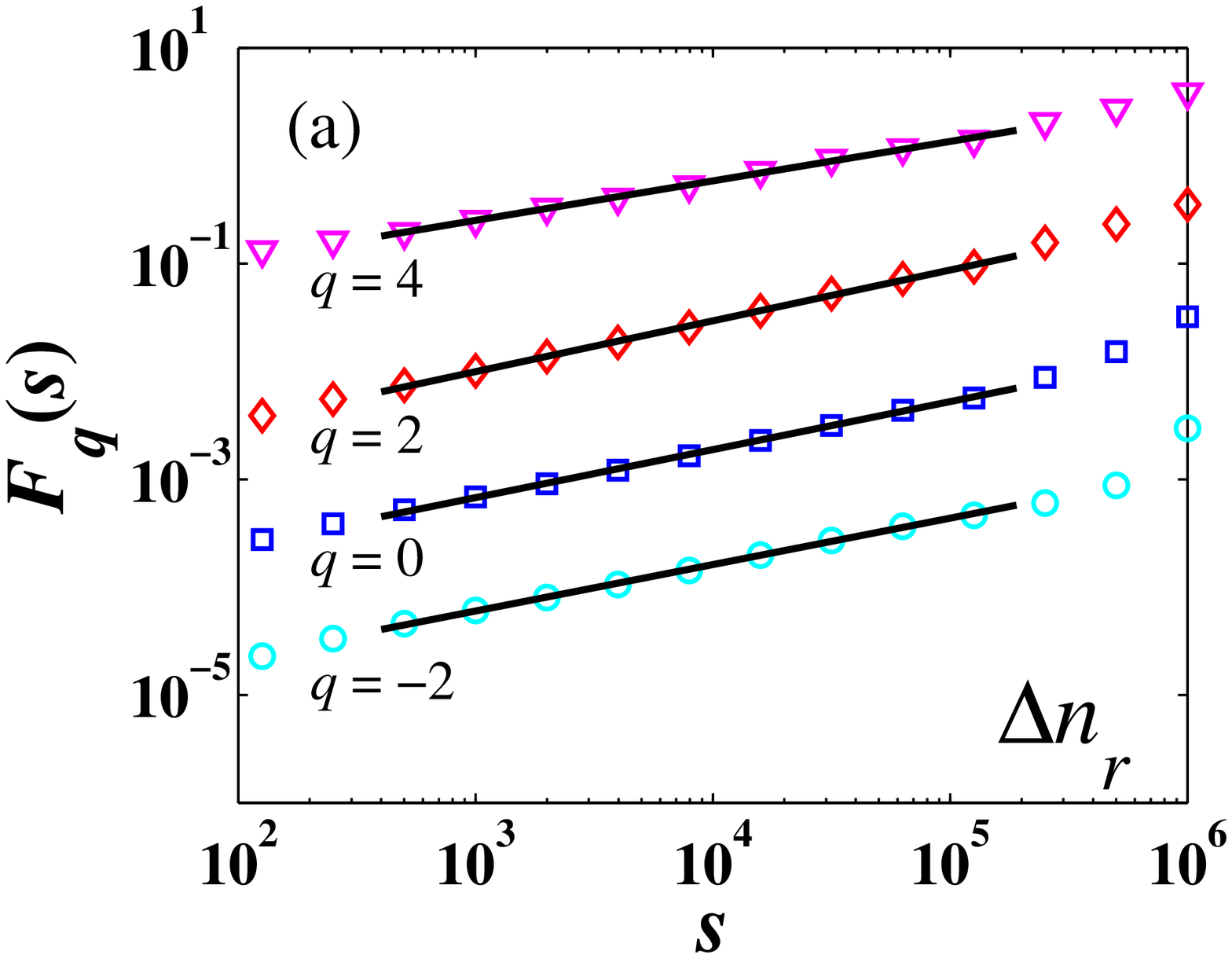}
\includegraphics[width=8cm]{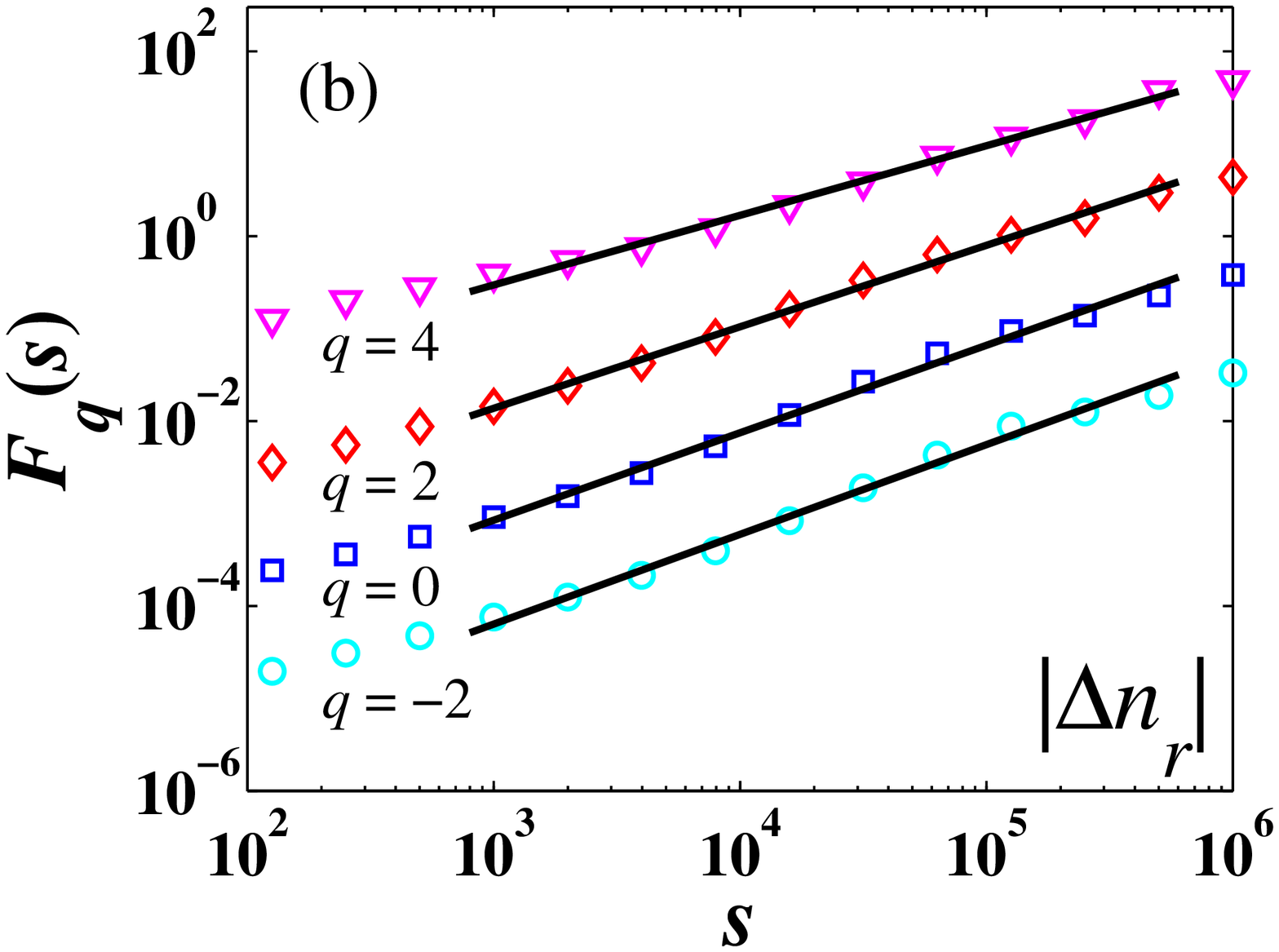}
\caption{(Color online) Dependence of the overall fluctuation
functions $F_q(s)$ with respect to the box size $s$ for $q = -2$, 0,
2 and 4 in log-log coordinates for (a) $\Delta n_r$ and (b) $|\Delta
n_r|$.} \label{Fig:dnr:MFDFA:Fs}
\end{figure}

The overall detrended fluctuations $F_q(s)$ are plotted as a
function of $s$ in log-log scales in Fig.~\ref{Fig:dnr:MFDFA:Fs} for
$\Delta n_r$ and $|\Delta n_r|$ and different orders $q$. The
power-law relation (\ref{Eq:scaling}) is verified for all the
curves, with the scaling ranges wider than two orders of magnitude.
An anomalous feature is observed in Fig.~\ref{Fig:dnr:MFDFA:Fs}(a)
for $\Delta n_r$ that the $F_{-2}(s)$ curve is flatter than the
$F_0(s)$ curve. It means that $h(-2)<h(0)$, which is not common even
for nonconservative quantities in multifractal analysis. This
anomaly is not observed  in Fig.~\ref{Fig:dnr:MFDFA:Fs}(b) for
$|\Delta n_r|$

The scaling exponent function $\tau(q)$, which is used to reveal the
multifractality in the standard multifractal formalism based on
partition function, can be obtained numerically as follows:
\begin{equation}
 \tau(q) = qh(q) - D_f,
  \label{Eq:scalingfunction}
\end{equation}
where $D_f$ is the fractal dimension of the geometric support of the
multifractal measure (in the current case we have $D_f = 1$). The
local singularity exponent $\alpha$ and its spectrum $f(\alpha)$ are
related to $\tau(q)$ through the Legendre transformation
\cite{Halsey-Jensen-Kadanoff-Procaccia-Shraiman-1986-PRA},
\begin{equation}
\left\{ \begin{aligned}
         \alpha &= {\rm{d}}\tau(q)/{\rm{d}}q \\
                  f(\alpha)&=q \alpha -\tau(q)
        \end{aligned} \right.~.
\label{Eq:alphaf}
\end{equation}
Since the size of each time series is finite, the estimate of
$F_q(s)$ will fluctuate remarkably for large values of $|q|$,
especially for large $s$.

\begin{figure}[htb]
\centering
\includegraphics[width=5.3cm]{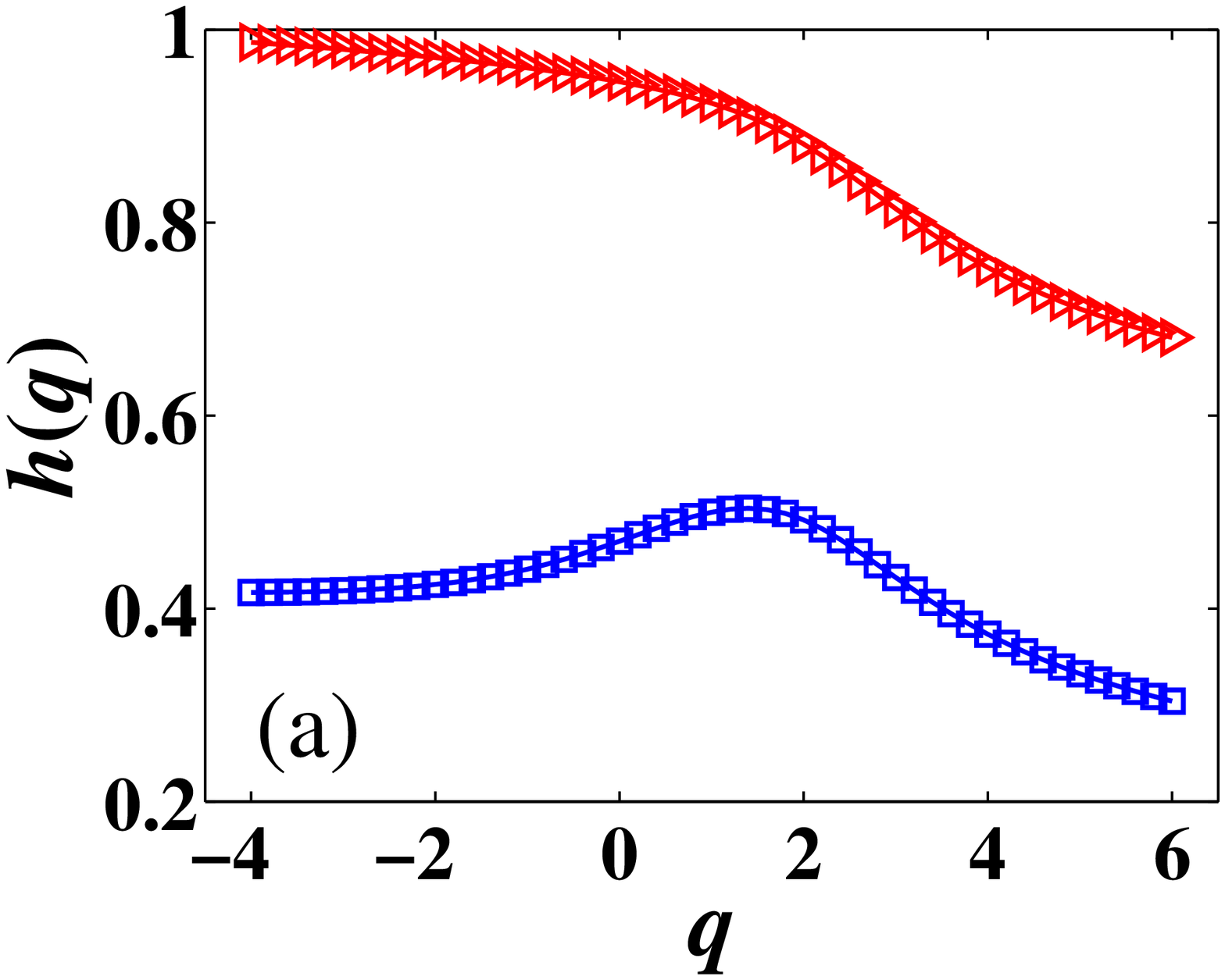}
\includegraphics[width=5.3cm]{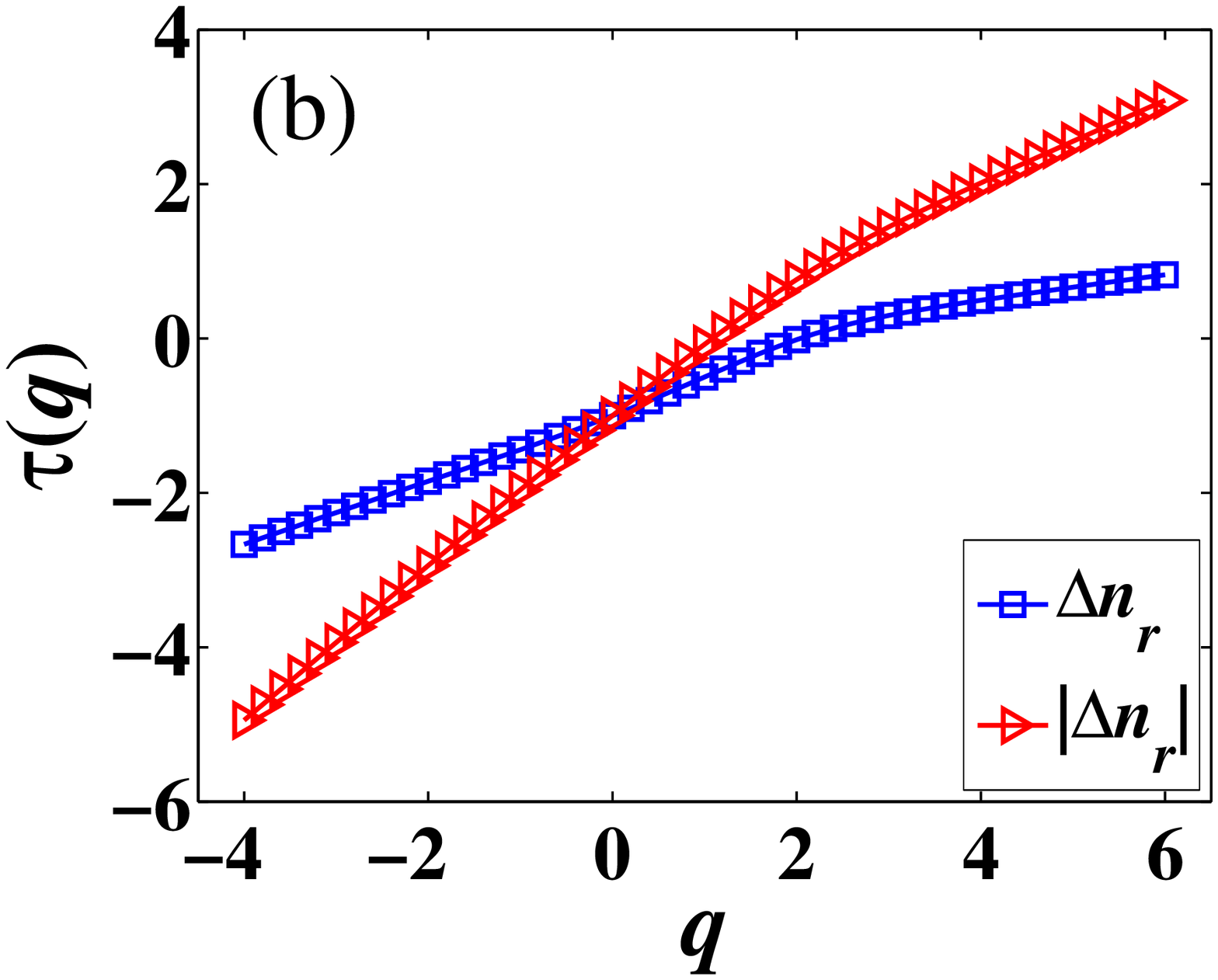}
\includegraphics[width=5.3cm]{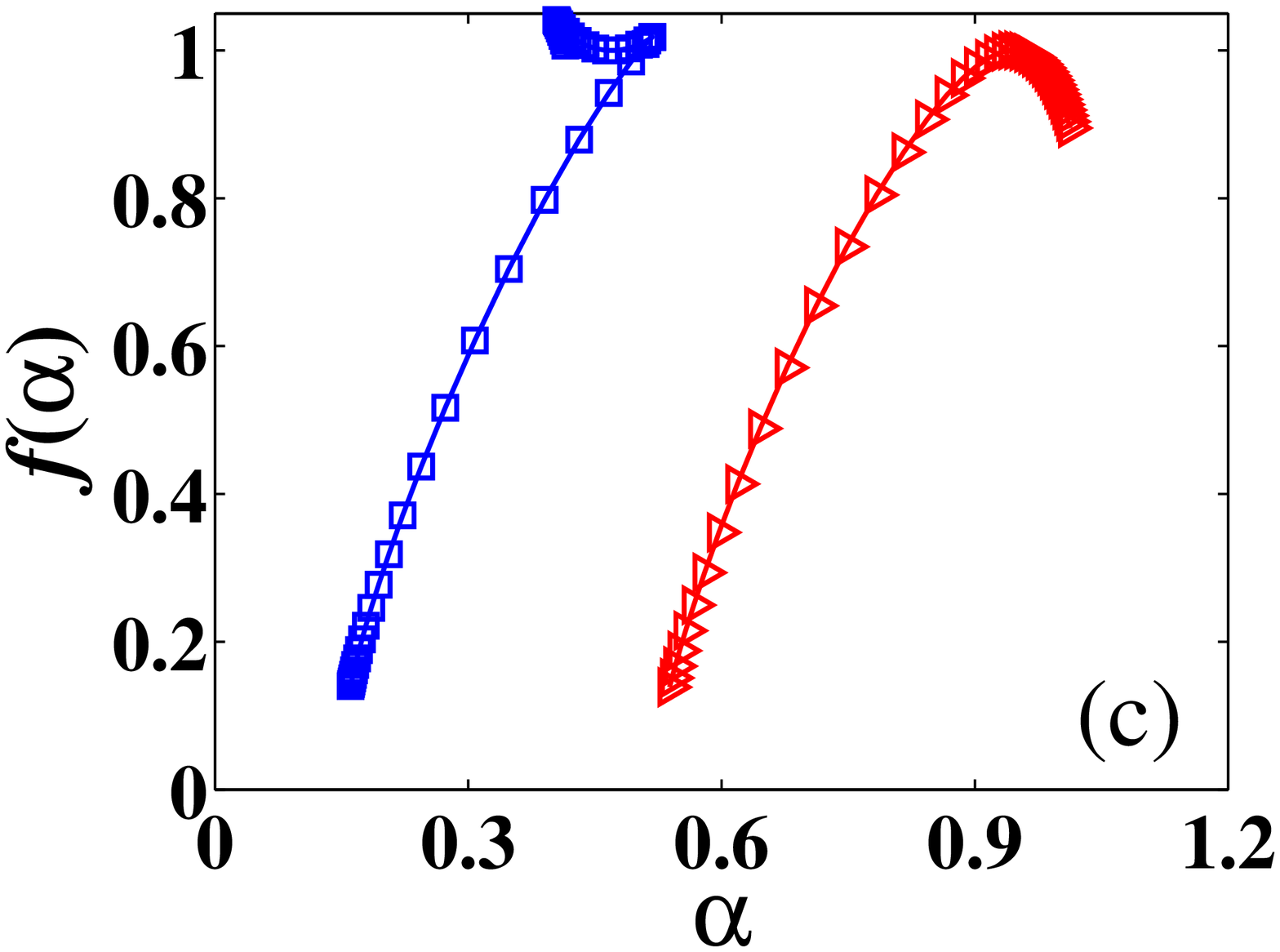}
\caption{(Color online) Multifractal analysis of $\Delta n_r$
($\square$) and $|\Delta n_r|$ ($\vartriangleright$). Shown are the
generalized Hurst indexes $h(q)$ (a), the mass exponents $\tau(q)$
(b), and the multifractal spectra $f(\alpha)$ (c).} \label{Fig:MF}
\end{figure}

Fig.~\ref{Fig:MF}(a) illustrates the generalized Hurst indexes
$h(q)$ as a function of $q$ for $\Delta n_r$ and $|\Delta n_r|$. We
find that the $h(q)$ function for $|\Delta n_r|$ decreases
monotonically, while that for $\Delta n_r$ increases in the left
part and decreases in the right part as indicated by
Fig.~\ref{Fig:dnr:MFDFA:Fs}(a). Similar phenomenon is rare with only
a few examples \cite{Jiang-Chen-Zhou-2009-PA}. Fig.~\ref{Fig:MF}(b)
shows the corresponding scaling exponent functions $\tau(q)$. The
nonlinearity in $\tau(q)$ is a hallmark for the presence of
multifractality in the time series. Fig.~\ref{Fig:MF}(c) presents
the two multifractal singularity spectra $f(\alpha)$ for the two
time series. Again, the spectrum $f(\alpha)$ for $\Delta n_r$
exhibits abnormal behavior.

\section{Conclusion}
\label{S1:Conclusion}

In summary, we have investigated the statistical properties of the
time series of instant number of online avatars in a massive
multiplayer online role-playing game. Spectral analysis shows that
the online avatar number exhibits one-day periodic behavior and
clear intraday pattern. On the contrary, our analysis suggests that
the maintaining of server should be scheduled before 7:30 a.m.
rather than in the wee hours. We also found that the fluctuations of
the online avatar numbers do not follow a Gaussian distribution.
Instead, the distribution is leptokurtic and fat-tailed. A maximum
likelihood method based on the Kolmogorov-Smirnov statistic shows
that the distribution has power-law tails. We also employed the
(multifractal) detrended fluctuation analysis to investigate the
memory effect of the increments of online avatar numbers after
removing the intraday pattern and the associated absolute values. We
found that the increments do not possess temporal correlation while
the absolute increments are long-term correlated. In addition, both
time series exhibit multifractal nature.

\bigskip
{\textbf{Acknowledgments:}}

This work was partly supported by the Program for New Century
Excellent Talents in University (Grant No. NCET-07-0288).

\bibliography{E:/Papers/Auxiliary/Bibliography}

\end{document}